\documentclass[aps,prmaterials,twocolumn,showpacs,amsmath,amssymb,superscriptaddress,floatfix]{revtex4-1}

\usepackage{color}         
\usepackage{graphics}      
\usepackage[pdftex]{graphicx}      
\usepackage{bm}
\usepackage{natbib}            
\usepackage[colorlinks,plainpages=false,linkcolor=blue,urlcolor=blue,citecolor=blue,pdfpagemode=UseNone,pdfstartview=FitBH]{hyperref}
\begin{document}

\title{Understanding the lattice thermal conductivity of SrTiO$_3$ from an \emph{ab initio} perspective}

\author{Adolfo O. Fumega}
  \email{adolfo.otero.fumega@usc.es}
 \affiliation{Departamento de F\'{i}sica Aplicada,
  Universidade de Santiago de Compostela, E-15782 Campus Sur s/n,
  Santiago de Compostela, Spain}
\affiliation{Instituto de Investigaci\'{o}ns Tecnol\'{o}xicas,
  Universidade de Santiago de Compostela, E-15782 Campus Sur s/n,
  Santiago de Compostela, Spain} 
 \affiliation{Department of Physics and Astronomy,
  University of Missouri, Columbia,
  Missouri 65211-7010, USA}
\author{Yuhao Fu}
\email{fuyuhaoy@gmail.com}
 \affiliation{Department of Physics and Astronomy,
  University of Missouri, Columbia,
  Missouri 65211-7010, USA}
\author{Victor Pardo}

\affiliation{Departamento de F\'{i}sica Aplicada,
  Universidade de Santiago de Compostela, E-15782 Campus Sur s/n,
  Santiago de Compostela, Spain}
\affiliation{Instituto de Investigaci\'{o}ns Tecnol\'{o}xicas,
  Universidade de Santiago de Compostela, E-15782 Campus Sur s/n,
  Santiago de Compostela, Spain} 
\author{David J. Singh}
  \email{singhdj@missouri.edu}
 \affiliation{Department of Physics and Astronomy,
  University of Missouri, Columbia,
  Missouri 65211-7010, USA}  
  
\begin{abstract} 

We present a detailed analysis of the structure dependence of the  lattice thermal conductivity of SrTiO$_3$. We have used both \emph{ab initio} Molecular Dynamic simulations and Density Functional Theory calculations to decouple the effect of different structural distortions on the thermal conductivity. We have identified two main mechanisms for tuning the thermal conductivity when a distortion is applied. First, the modification of the acoustic-modes energy dispersion when a change in the lattice parameters is imposed and second, the low energy polar modes. In particular and counterintuitively, we have found that an increase in the angle of the oxygen octahedral rotations increases the thermal conductivity due to its coupling to these polar modes.

\end{abstract}

\maketitle

\section{Introduction}

Complex oxides have become a major platform for discovering new fundamental physical properties of materials and also for their applications in devices. From the high-temperature superconducting cuprates\cite{norman_review_sc} to the perovskite photovoltaics \cite{grinberg2013perovskite}, understanding the underlying physical properties of complex oxides is key in condensed matter physics and its applications. The cubic perovskite structure has always been a paradigmatic example. In particular, SrTiO$_3$ (STO) serves as a template for all sorts of oxide-based nanostructures since many other oxides can be grown on top of it, but also because of its own intriguing physical properties. It is a quantum paraelectric material that undergoes a structural transition at 105 K \cite{doi:10.1143/JPSJ.23.546}. This is a tetragonal (low-temperature) to cubic (higher temperature) transition accompanied of an octahedral rotation existing in the low-temperature phase. It also becomes superconducting at 300 mK when doped \cite{sto_sc_1964}. Its nanostructures have shown all sorts of interesting physical properties: two-dimensional electron gas \cite{ohtomo_hwang_2004}, ferromagnetism \cite{brinkman2007magnetic}, superconductivity \cite{lao_sto_sc}, a tunable Rashba effect \cite{lao_sto_rashba}, etc. The influence of various sorts of strain have been also thoroughly studied, in particular the appearance of a ferroelectric-like phase on the tensile strain limit has been predicted \cite{Haeni2004} but not fully demonstrated experimentally so far. 
Due to its potential as an energy harvesting material, the thermoelectric properties of STO have been profusely studied \cite{ohta_sto_seebeck_nmat_2007,sto_sarantopoulos}.
However, its thermal conductivity is probably the least studied factor from a computational point of view \cite{LINDSAY2018106}.

Different strategies have been proposed to tune the thermal conductivity of STO: nanostructuring \cite{BUSCAGLIA2014307,LI2016107}, application of external electric fields \cite{PhysRevMaterials.3.044404}, substitution of different kind of cations \cite{Wang2013,C4RA06871H,kappa_defects,PhysRevLett.120.125901,Yaremchenko2015,CHEN201988,Dawson2014}, oxygen vacancies \cite{Zhang_2017,oxygen_vac,sto_kappa_stoichiometry}, etc. Apart from that, a Poiseulle flow of phonons was recently identified on STO at low temperatures \cite{PhysRevLett.120.125901}.
Yet despite all this, there are interesting fundamental questions one can ask about heat transport in STO: what are the main mechanisms that determine its thermal conductivity? How can we tune them by modifying its structure?

In this work, we analyze from a theoretical point of view how different structural distortions affect the thermal conductivity of STO. First, we have identified which are the main features that determine its thermal conductivity. Then, we have seen how those get modified when a particular distortion is applied to the structure. We have performed this analysis in both the high temperature cubic phase and the low temperature tetragonal one. The results obtained in this study can serve as a platform for drawing a more general conclusion in complex oxides and their nanostructures, as well as a guidance to interpret experimental results.

\begin{figure*}
  \centering
  \includegraphics[width=1.0\textwidth]%
    {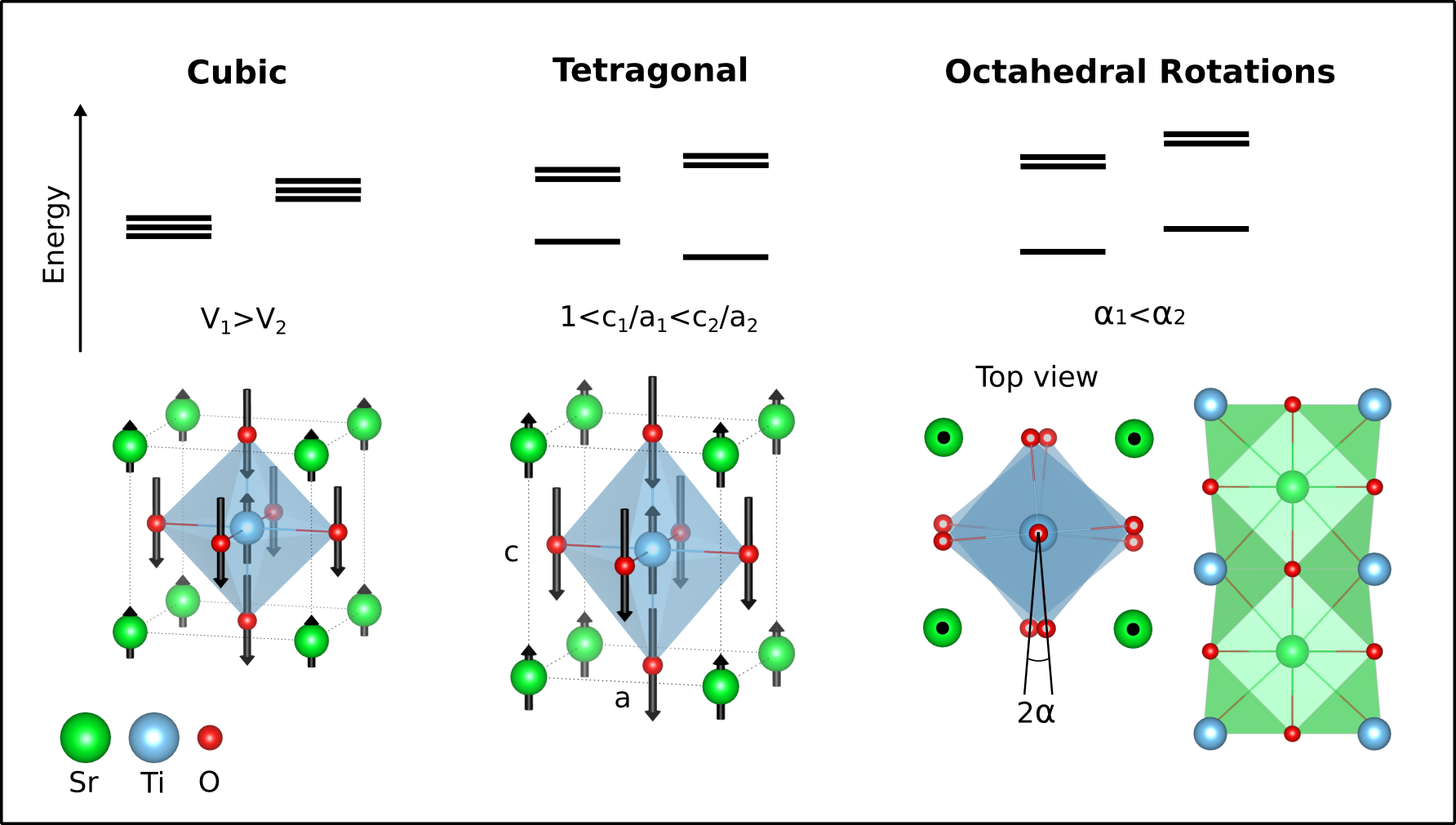}
     \caption{Energetic scheme of the polar modes that tune the thermal conductivity. Sr (Ti, O) atoms depicted in green (blue, red). Black arrows illustrate the atomic displacement in the polar mode. For the cubic phase, since the three lattice parameters are equal, the mode is triply degenerate. Decreasing the volume of the unit cell increases the frequency of the mode and hence the thermal conductivity is enhanced. The tetragonal distortion breaks the triple degeneracy. For $c/a>1$, the polar mode along the z-axis lowers its frequency due to the bond length increase between the Ti atom and the apical O atoms of the octahedra. Increasing the $c/a$ ratio increases the gap between xy and z-axis modes, thus reducing the thermal conductivity. The octahedral rotations keep the volume of the octahedra constant. However, the Sr-O bond length is reduced. This causes an increase on the frequency of the mode producing an increase in the thermal conductivity.}\label{scheme_polar}
\end{figure*}

\section{Computational methods}

We report first principles calculations of the lattice thermal conductivity ($\kappa$) for STO in its cubic and tetragonal phases by iteratively solving the linearized Boltzmann-Peierls transport equation of phonons with the ShengBTE package \cite{ShengBTE_2014}.  Converged phonon momenta $q$ meshes of $15\times15\times15$ were used for solving the transport equation. 
In order to calculate $\kappa$, stable structures (i.e., phonon band structures without imaginary frequencies) are required. Thus, two different methods were used to compute each phase. We have also used two different codes for each method. We have checked that both codes provide compatible results.

For the high temperature $Pm\overline{3}m$ cubic phase which is unstable at 0 K, \emph{ab initio} Molecular Dynamic (aiMD) calculations were performed as implemented in the {\sc VASP} code \cite{VASP} in order to stabilize the cubic phase that occurs at high temperature. The simulations were carried out in a $2\times2\times2$ supercell (which was checked to be big enough to converge the thermal conductivity) in the canonical ensemble at 300 K during 20000 steps of 1 fs following equilibration. 
The generalized gradient approximation (GGA) in Perdew-Burke-Ernzerhof (PBE) scheme \cite{PBE} 
was used for the exchange-correlation term within the projector augmented wave (PAW) method. 
The choice of the functional was based on previous studies on lattice dynamics of STO \cite{Feng_2015}. We have also checked that the results that we have obtained with it are in agreement with experiments. The tolerance in the forces was 10$^{-4}$ eV$/$\AA, and the cutoff energy for the planewave basis set was 410 eV. 
We have used the temperature-dependent effective potential (TDEP) method \cite{TDEP} to extract the effective interatomic force constants (IFCs) that best describe the anharmonic Born-Oppenheimer potential energy surface at a given temperature from the set of supercells with  displacements generated in the aiMD.  

For the low-temperature $I4/mcm$ tetragonal phase, we have performed first principles Density Functional Theory (DFT) calculations \cite{HK,KS} with the PAW method as implemented in the {\sc quantum espresso} code \cite{quantum_espresso}. The GGA in the revised PBE scheme \cite{PBEsol} was used for the exchange-correlation term. The harmonic IFCs were calculated using density-functional perturbation theory (DFPT) \cite{phonons_dfpt,quantum_espresso}. The third-order anharmonic IFCs were computed using the real-space supercell approach \cite{ShengBTE_2014} with a $3\times3\times3$ supercell. The single cell was the 10-atom primitive cell that corresponds to the low-temperature $I4/mcm$ tetragonal phase. All the calculations were performed in a converged \emph{k}-mesh with a plane-wave energy cutoff of 60 Ry for the kinetic energy and 240 Ry for the charge density. The longitudinal optical-transverse optical (LO-TO) splitting was included using the dielectric constant and Born effective charges calculated from linear response DFPT.

\section{Results and discussion}

Figure \ref{scheme_polar} shows the kind of structures we will be considering in this work, based on the perovskite structure. We will analyze first the high temperature phase. We will identify the main sources that determine the thermal conductivity of STO. Then, we will proceed to analyze the low temperature phase and how different distortions modify the thermal conductivity. 

\subsection{The high temperature $Pm\overline{3}m$ phase}

\begin{figure}[!h]
  \centering
  \includegraphics[width=\columnwidth]%
    {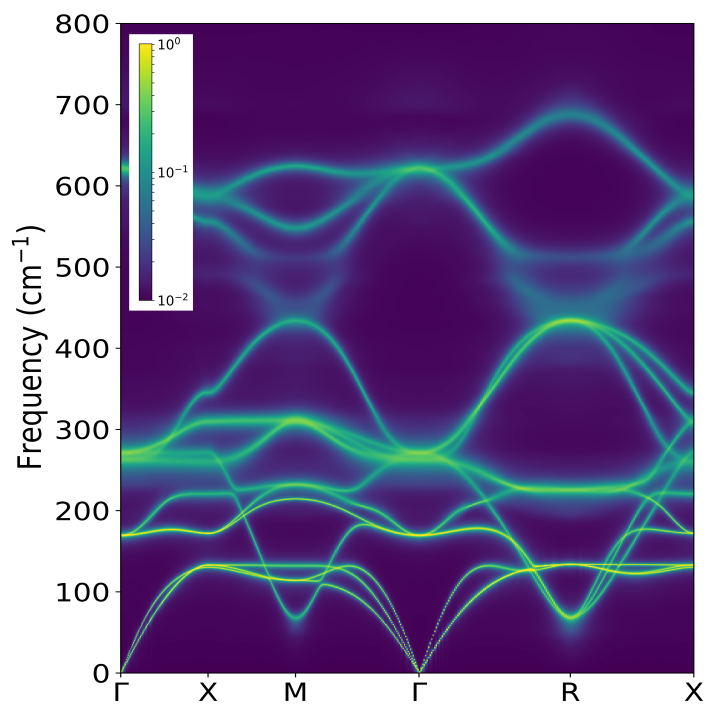}
     \caption{Phonon lineshape in the $Pm\overline{3}m$ phase at 300 K. The intensity of the bands represents the likelihood of exciting a particular phonon. The high peaks of the intensity are found at the acoustic bands and at the polar modes around 200 cm$^{-1}$.}\label{bandscub}
\end{figure}

First, we start analyzing the lattice thermal conductivity in the cubic phase. This phase is the stable one at high temperatures. We have imposed the experimental lattice parameter of cubic STO, $a_0=3.905$ \AA \cite{sto_sarantopoulos}. The calculation of the phonon band structure of this phase from DFPT produces imaginary frequencies at the R point of the Brillouin zone as may be expected. These unstable modes are related to the octahedral rotations that appear at low temperatures \cite{PhysRevB.83.134108}.  Figure \ref{bandscub} shows the phonon dispersion for cubic STO computed using aiMD. As we have already mentioned, this method allows to stabilize the high temperature phase, permitting us to compute the thermal conductivity in the cubic phase. Its calculated value is 13.4 W/mK which is in reasonable agreement with the one reported by experiments (11 (W/mK) at 300 K)\cite{APL_eric}. The small overestimation in our calculation may perhaps be due to the lack of impurities and boundary scattering in our calculations as opposed to the experimental case.

\begin{figure}[!h]
  \centering
  \includegraphics[width=\columnwidth]%
    {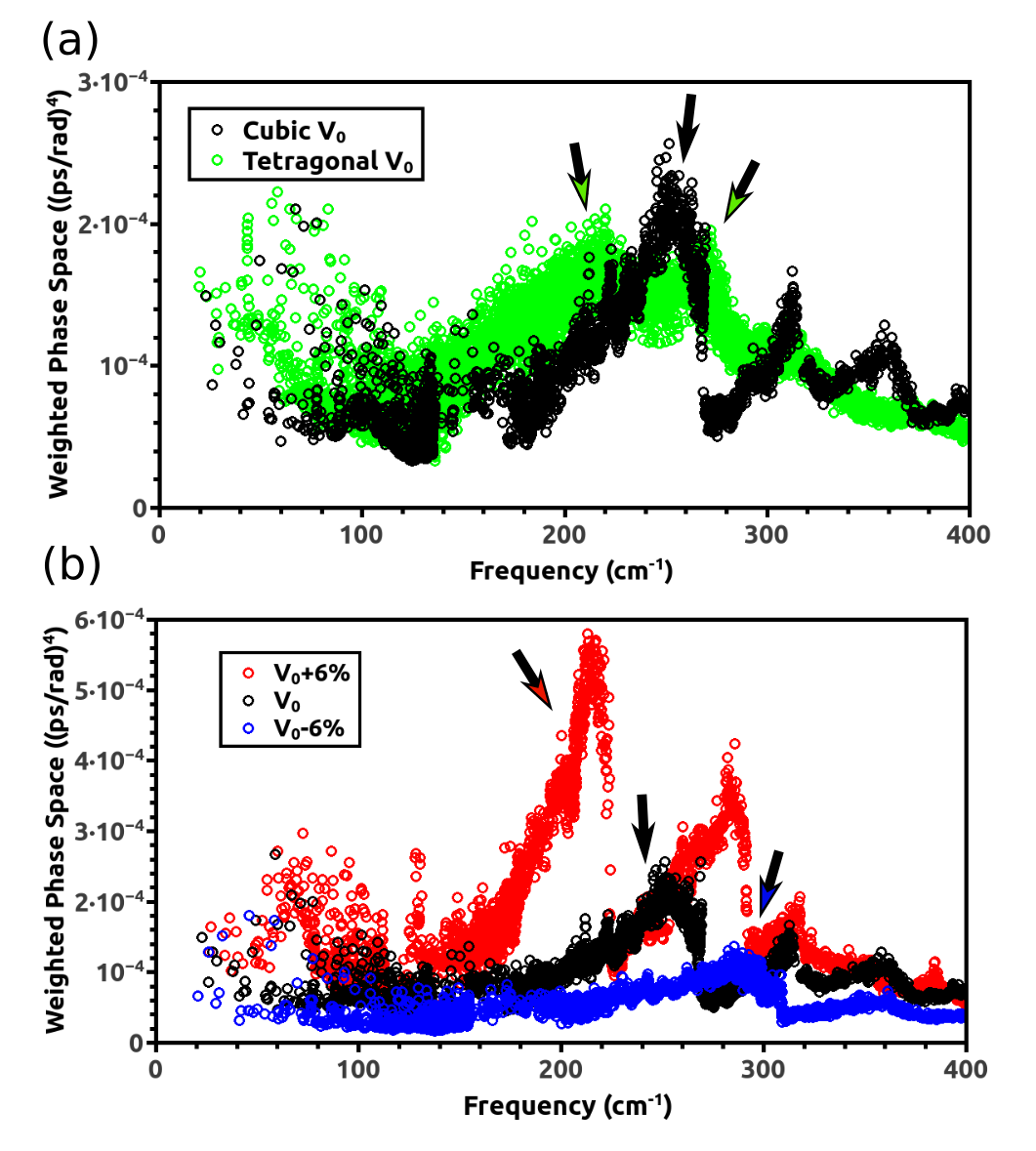}
     \caption{Weighted phase space for the high temperature phase at 300 K as a function of frequency. An increase is related to an increase of the anharmonic scattering rate, thus decreasing the thermal conductivity. The arrows indicate the peaks associated to the polar modes that tune the thermal conductivity. (a) Comparison between cubic (black) and tetragonal (green) structures. (b) Evolution with the unit cell volume. }\label{w3p_300}
\end{figure}

In order to identify the modes that contribute most to the thermal transport, we have computed and plotted in the phonon band structure with the so called lineshape (Fig. \ref{bandscub}). This lineshape is nothing else but the one-phonon neutron scattering cross section that can be measured in inelastic neutron experiments. The frequency axis is the probing energy and the intensity represents the likelihood of exciting a phonon at a given energy and momentum. The lineshape is related to the imaginary part of the phonon self-energy, which can be computed via the third order force constants \cite{PhysRev.128.2589, Cowley_1968, wallace1998thermodynamics}. The most intense and sharp peaks in the lineshape provide the most active phonon modes in the sense of the longest lifetimes. Therefore, they will be the ones that contribute most to the thermal transport. In the case of STO, we can see in Fig. \ref{bandscub}, that they correspond to the acoustic bands and to the triply degenerate polar mode depicted in Fig. \ref{scheme_polar} (the modes below 250 cm$^{-1}$). The order of magnitude of the STO thermal conductivity is given by the dispersion of the acoustic bands, since these modes are the main heat carriers, but also tuned by the polar modes. The mean group velocity associated to the acoustic modes at 300 K that we have obtained is 5.24 km/s, which is in good agreement with previous experiments \cite{APL_eric}.

\begin{figure}[!h]
  \centering
  \includegraphics[width=\columnwidth]%
    {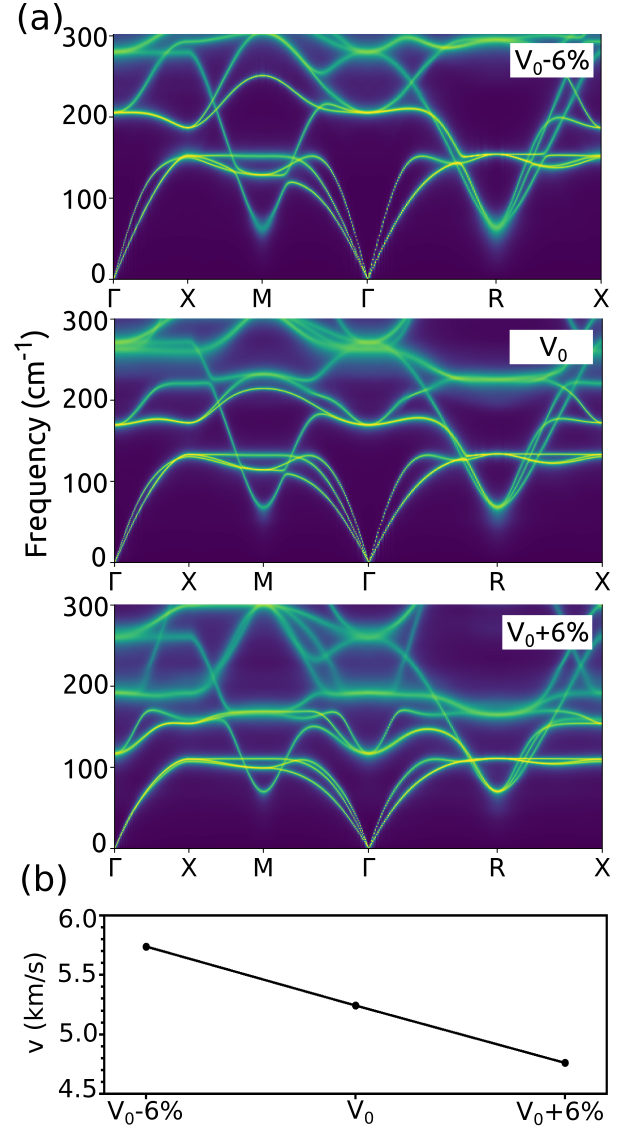}
     \caption{(a)Phonon lineshape for three different volumes in the $Pm\overline{3}m$ phase at 300 K. From top panel to bottom we present an increase in the volume of the unit cell, $V_{0}$ being the experimental unit cell volume. At the $\Gamma$ point and between $100-200$ cm$^{-1}$, we identify the triply degenerate polar mode.  Increasing the volume decreases the frequency of the polar mode. The band dispersion of the acoustic modes increases when reducing the volume.(b) Phonon mean group velocity of the acoustic bands as a function of the unit cell volume.}\label{bandscub_vol}
\end{figure}

In Fig. \ref{w3p_300} we have plotted the weighted phase space as defined in \cite{def_w3p}, it will help us to identify the main sources of scattering that tune the thermal conductivity of STO. The weighted phase space provides the scattering rate as a function of frequency weighted by the harmonic frequencies. Therefore, it is  very sensitive to the phonon spectra and hence can be used to identify the modes responsible for the stronger scattering processes. It also helps to analyze how the scattering rate is modified when different distortions, that change the phonon spectra, are applied to the structure.
In the case of cubic STO at 300 K, we can see in Fig. \ref{w3p_300}a (black dots) that the weighted phase space has a peak around 250 cm$^{-1}$. At this frequency, this peak is associated to the scattering processes of the polar modes. So, we can conclude that the polar modes shown in Fig. \ref{scheme_polar} are the main source of scattering in STO. This triply degenerate polar mode was previously identified, using other methods, as one of the main sources of phonon-phonon scattering \cite{Feng_2015, PhysRevMaterials.3.044404}. Apart from that, in a recent work, this mode was also reported to be electron-phonon active \cite{PhysRevLett.121.226603}. 

The subsequent discussions will be based on how the acoustic and polar modes evolve under different distortions. Distortions that shift the heat carrying modes may be expected to affect the thermal conductivity via a change in the phonon dispersion of the acoustic modes and also in the scattering rate. In particular shifting the polar modes to lower frequency may be expected to more strongly scatter acoustic phonons lowering the thermal conductivity.

\begin{figure}[!h]
  \centering
  \includegraphics[width=\columnwidth]%
    {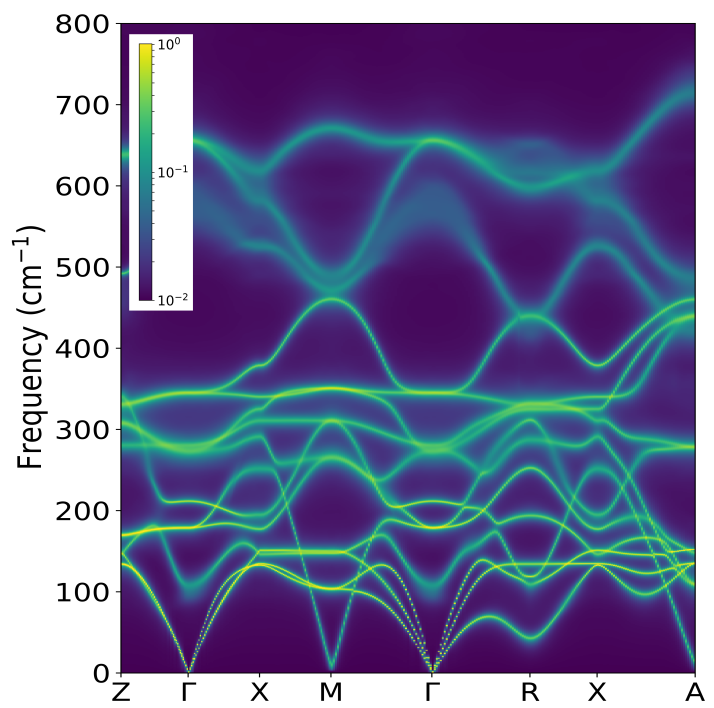}
     \caption{Phonon lineshape for a tetragonal distortion of the cubic phase at 300 K ($c/a=1.06$). The triple degeneracy of the polar mode is broken due to this distortion as explained in Fig. \ref{scheme_polar}. There is a softening of the polar mode along the z-axis.}\label{bandstet}
\end{figure}

Figure \ref{bandscub_vol}a shows the lineshape for three different unit cell volumes at 300 K for a cubic structure. 
We can observe that a decrease in the volume of the unit cell increases the dispersion of the acoustic bands. This is translated into an increase of the mean group velocity of the acoustic bands (Fig. \ref{bandscub_vol}b). A reduction of the unit cell volume also increases the frequency of the triply degenerate polar mode due to the reduction of the Ti-O bond length. This increase in the frequency (depicted in Fig. \ref{scheme_polar}) reduces the phonon scattering, as can be seen in Fig. \ref{w3p_300}b, where a clear reduction of the phase space peak occurs by reducing the unit cell volume. Hence, both effects on the acoustic and polar modes produce an increase in the thermal conductivity. The calculated values are: 7.6, 13.4 and 18.0 W/mK for the largest unit cell volume to the smallest one, respectively. This increase of the thermal conductivity when the volume of the unit cell is reduced is in agreement with previous experiments on other family of oxides that report an increase of $\kappa$ when pressure is applied \cite{HASEGAWA2019229}. 

\begin{figure}[!h]
  \centering
  \includegraphics[width=\columnwidth]%
    {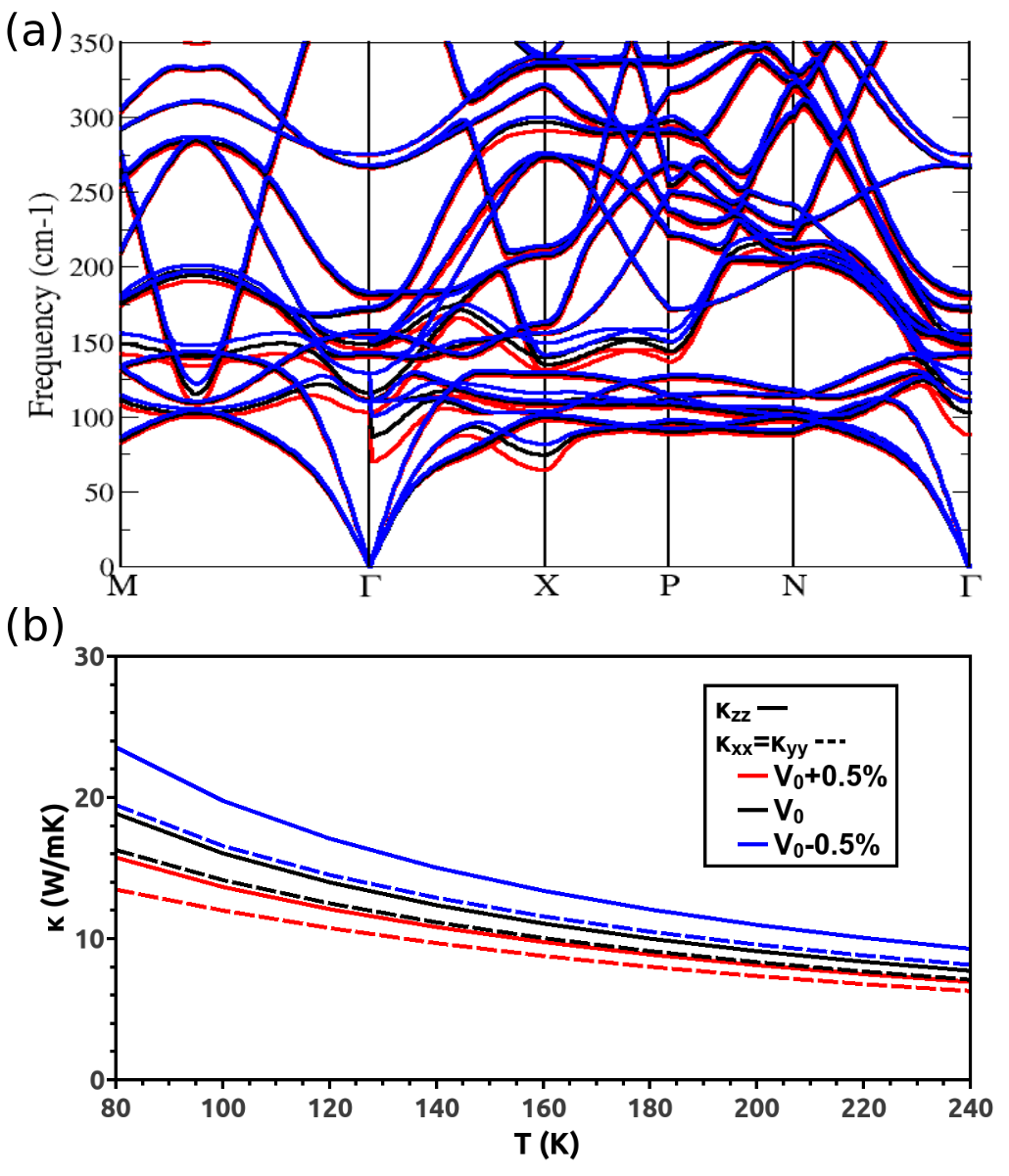}
     \caption{Results for the variation of the unit cell volume in the $I4/mcm$ phase. Octahedral rotation angle and $c/a$ ratio are kept constant. (a) Phonon band structure for three different volumes, in red the largest and in blue the smallest, $V_{0}$ is the optimized unit cell volume. An increase in the volume reduces the frequency of the polar modes and the phonon dispersion of the acoustic modes increases. (b) Thermal conductivity as a function of temperature. An increase in the volume reduces the thermal conductivity.}\label{kappa_vol}
\end{figure}

Figure \ref{bandstet} shows the lineshape computed at 300 K when a tetragonal distortion is applied. This distortion keeps the volume of the unit cell constant and equal to the experimental one, while the $c/a$ ratio was kept fixed at 1.06. We can see that the triple degeneracy of the polar mode is broken as explained in Fig. \ref{scheme_polar}. Since the $c$ lattice parameter becomes larger it will increase the bond length of the Ti with the apical O atoms of the octahedra, reducing the frequency of the mode along the z-axis as shown in Fig. \ref{bandstet}. Moreover, due to the increase of the $c$ lattice parameter and the decrease of $a$, the acoustic modes change their dispersion as compared to the cubic case.  
The calculated lattice thermal conductivity is $\kappa_{xx}=\kappa_{yy}=10.4$, $\kappa_{zz}=8.0$ W/mK.
We see that the tetragonal distortion breaks the triple degeneracy of the $\kappa$-tensor and reduces the thermal conductivity at 300 K. The degeneracy break of the polar mode is the main responsible for decreasing the thermal conductivity. In Fig. \ref{w3p_300}a we can see how the peak associated to the polar modes is divided in two peaks (green dots) when a tetragonal distortion is applied. The difference that we observe between cubic and tetragonal thermal conductivity would be diminished taking into account the volume reduction that occurs at the cubic to tetragonal phase transition, which produces an increase on the thermal conductivity. Furthermore, we will see in the next subsection that including the octahedral rotations in the tetragonal phase will increase the thermal conductivity. Thus, the difference obtained between both phases would be reduced by including these effects.

\begin{figure}[!h]
  \centering
  \includegraphics[width=\columnwidth]%
    {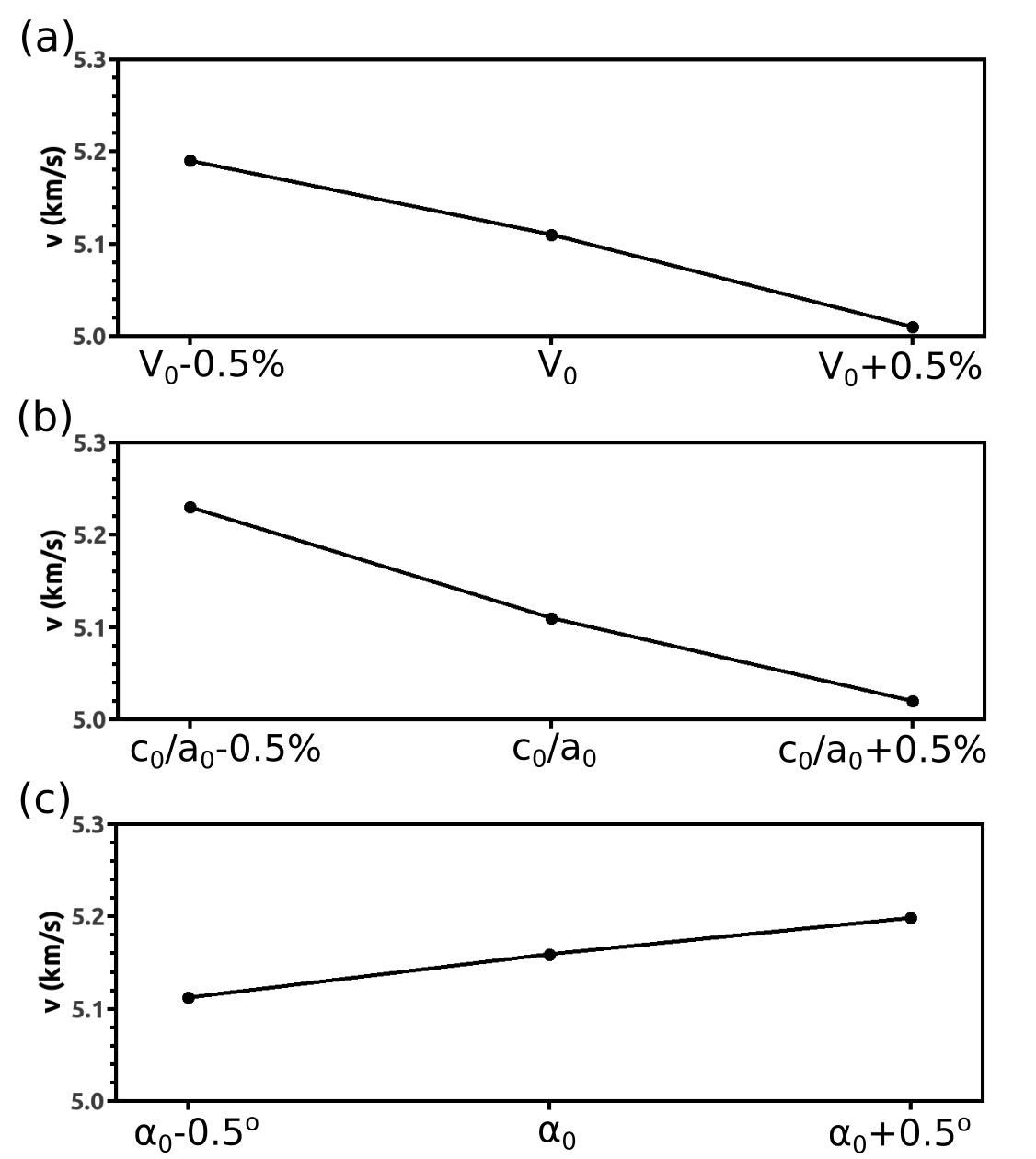}
     \caption{Mean group velocity of the acoustic modes for the low temperature phase. Its increase produces an increase of the thermal conductivity. (a) Evolution with the unit cell volume, (b) $c/a$ ratio and (c) octahedral rotation angle.}\label{dft_vel}
\end{figure}

\subsection{The low temperature $I4/mcm$ phase}

Below 105 K, STO undergoes a phase transition from the cubic $Pm\overline{3}m$ space group to the tetragonal $I4/mcm$ \cite{doi:10.1143/JPSJ.23.546}. This tetragonal phase is characterized by the stretching of the unit cell and also by a TiO6$_6$ octahedra antiphase rotation along the c axis \cite{RevModPhys.46.83}. The structure of this phase is depicted in Fig. \ref{scheme_polar}. We have optimized the lattice parameters and the atomic positions. We have then applied different structural distortions and study how they affect the phonon dispersion and also the thermal conductivity. We will see that, as obtained for the high temperature cubic phase, the polar modes play a fundamental role in the heat transport. The optimized structure that we have obtained has $a_0=3.868$ \AA, $c_0=3.935$ \AA, lattice parameters and $\alpha_0=6.6^{\circ}$ octahedral rotation angle in agreement with previous calculations \cite{PhysRevB.83.134108}. Both $c_0/a_0$ ratio and $\alpha_0$ are overestimated by GGA compared to experiment values \cite{PhysRevB.60.2961}. Since the optimized structure is a stable structure at 0 K, we can perform the study using DFT calculations. Small distortions to this structure can be applied without introducing imaginary frequencies in the phonon spectrum that would invalidate our analysis. In this subsection we will study the effect on the spectrum and the thermal conductivity of unit cell volume, $c/a$ ratio and octahedral rotation angle. We will keep two of these parameters constant while varying the third one to try to decouple their effects. 

\begin{figure}[!h]
  \centering
  \includegraphics[width=\columnwidth]%
    {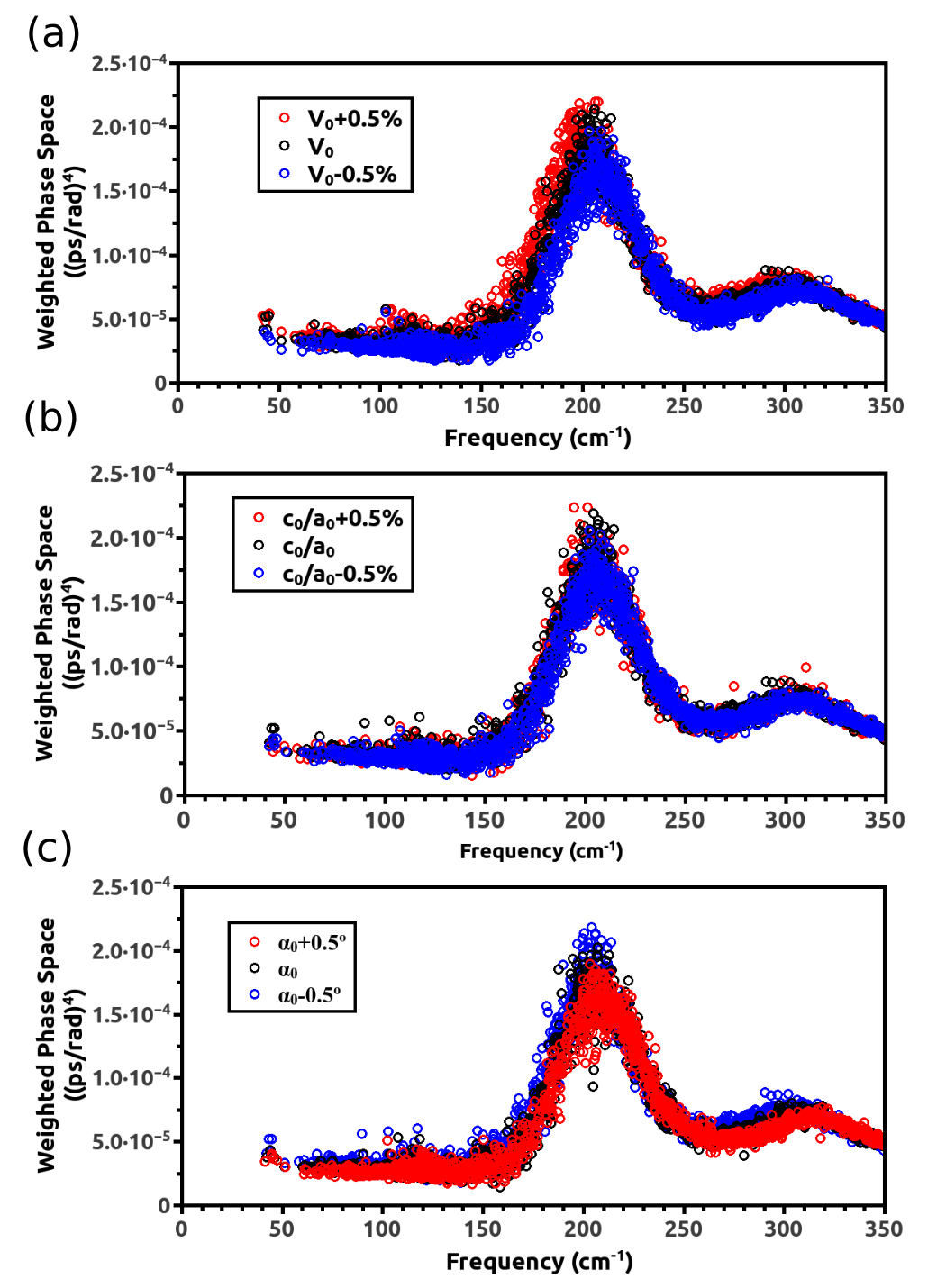}
     \caption{Weighted phase space for the low temperature phase at 80 K as a function of frequency. The high peak around 200 cm$^{-1}$ is associated to the polar modes. Its increase produces an increase of the anharmonic scattering rate and hence a reduction of the thermal conductivity (a) evolution with the unit cell volume, (b) $c/a$ ratio and (c) octahedral rotation angle.}\label{dft_w3p}
\end{figure}

Note that we will plot the thermal conductivity of this low temperature phase for temperatures above 105 K, the temperature of the structural phase transition. We do this to make visually clear the variations in $\kappa$ with each distortion. Moreover, below 80 K the thermal conductivity is not plotted since it is dominated by the sample dependent boundary scattering. Therefore, the realistic temperature range for the computed $\kappa$ is between 80 and 105 K. 

\begin{figure}[!h]
  \centering
  \includegraphics[width=\columnwidth]%
    {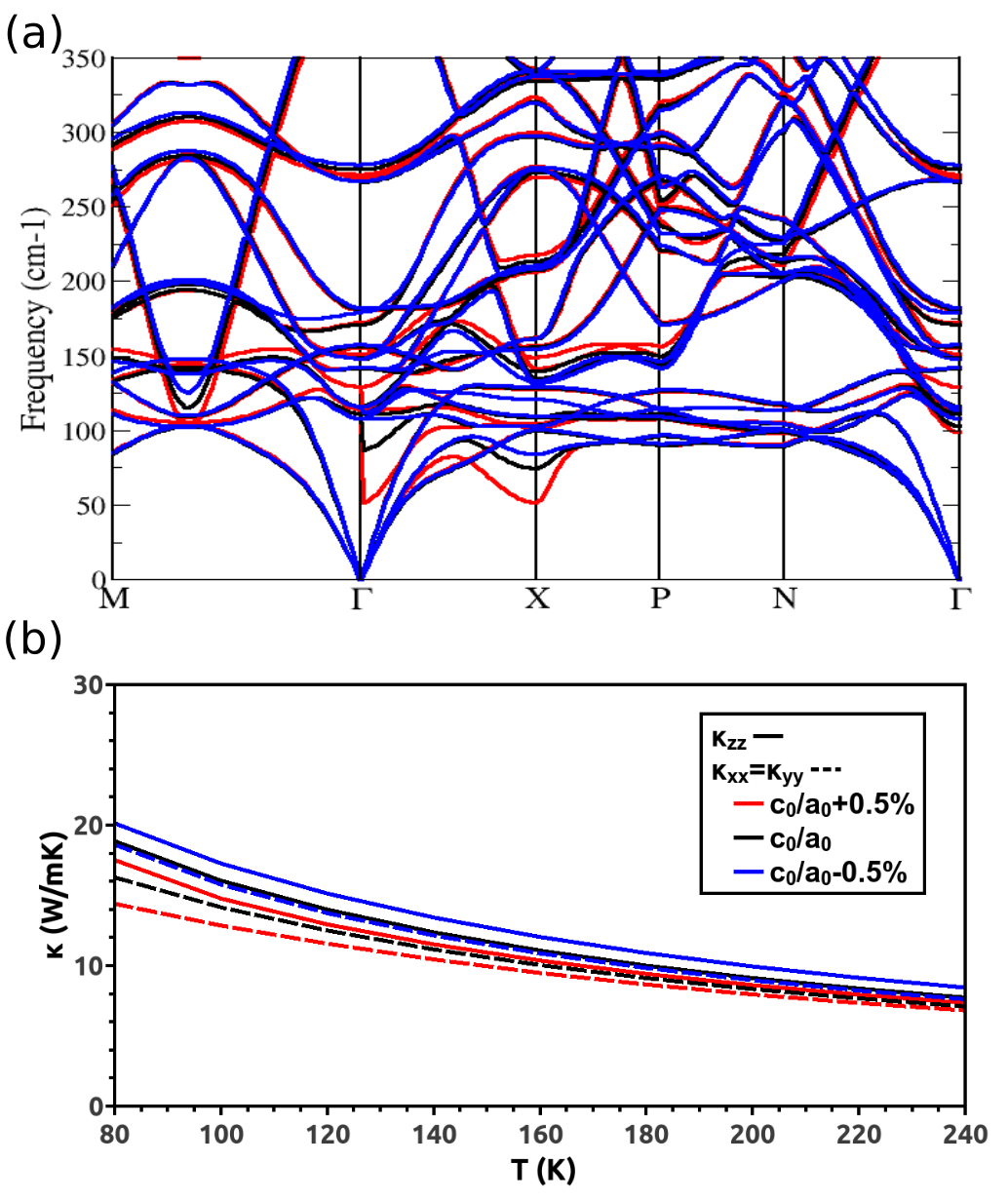}
     \caption{Results for variation of the $c/a$ ratio in the $I4/mcm$ phase. Unit cell volume and octahedral rotation angle are kept constant. (a) Phonon band structure for three different $c/a$ ratios, in red the largest and in blue the smallest, $c_{0}/a_{0}$ is the optimized $c/a$ ratio. The acoustic bands undergo a softening, mostly at the X point, when the $c/a$ ratio is increased. At the $\Gamma$ point, we see that an increase of the $c/a$ ratio increases the gap between the xy and z-axis polar modes. (b) Thermal conductivity as a function of temperature. Increasing the $c/a$ ratio lowers the thermal conductivity.}\label{kappa_ac}
\end{figure}

Figure \ref{kappa_vol}a shows the phonon dispersion for three different volumes of the unit cell. We can see that a reduction of the volume increases the dispersion of the acoustic bands, thus increasing the group velocity of the compound (Fig. \ref{dft_vel}a). We can also see that the polar modes increase their frequency when the volume is reduced. This causes a reduction of the phonon-phonon interaction.  Figure \ref{dft_w3p}a shows how the peak of the weighted phase space associated to the polar modes decreases when the volume is reduced. These two consequences in the phonon spectrum act in a positive way increasing the thermal conductivity when reducing the volume (see Fig. \ref{kappa_vol}b). This trend is in agreement with the result obtained for the high temperature cubic phase. The thermal conductivity at 100 K for the optimized structure is $\kappa_{zz}=16.1$ $W/mK$ and $\kappa_{xx}=\kappa_{yy}=14.2$ $W/mK$. Considering that $V_0$ is overestimated by GGA, these are in pretty good agreement with the experimental value \cite{APL_eric} and also with previous calculations \cite{PhysRevMaterials.3.044404}. We see that a decrease in $1\%$ in the unit cell volume produces an increase of around $30\%$ on the thermal conductivity.

\begin{figure}[!h]
  \centering
  \includegraphics[width=\columnwidth]%
    {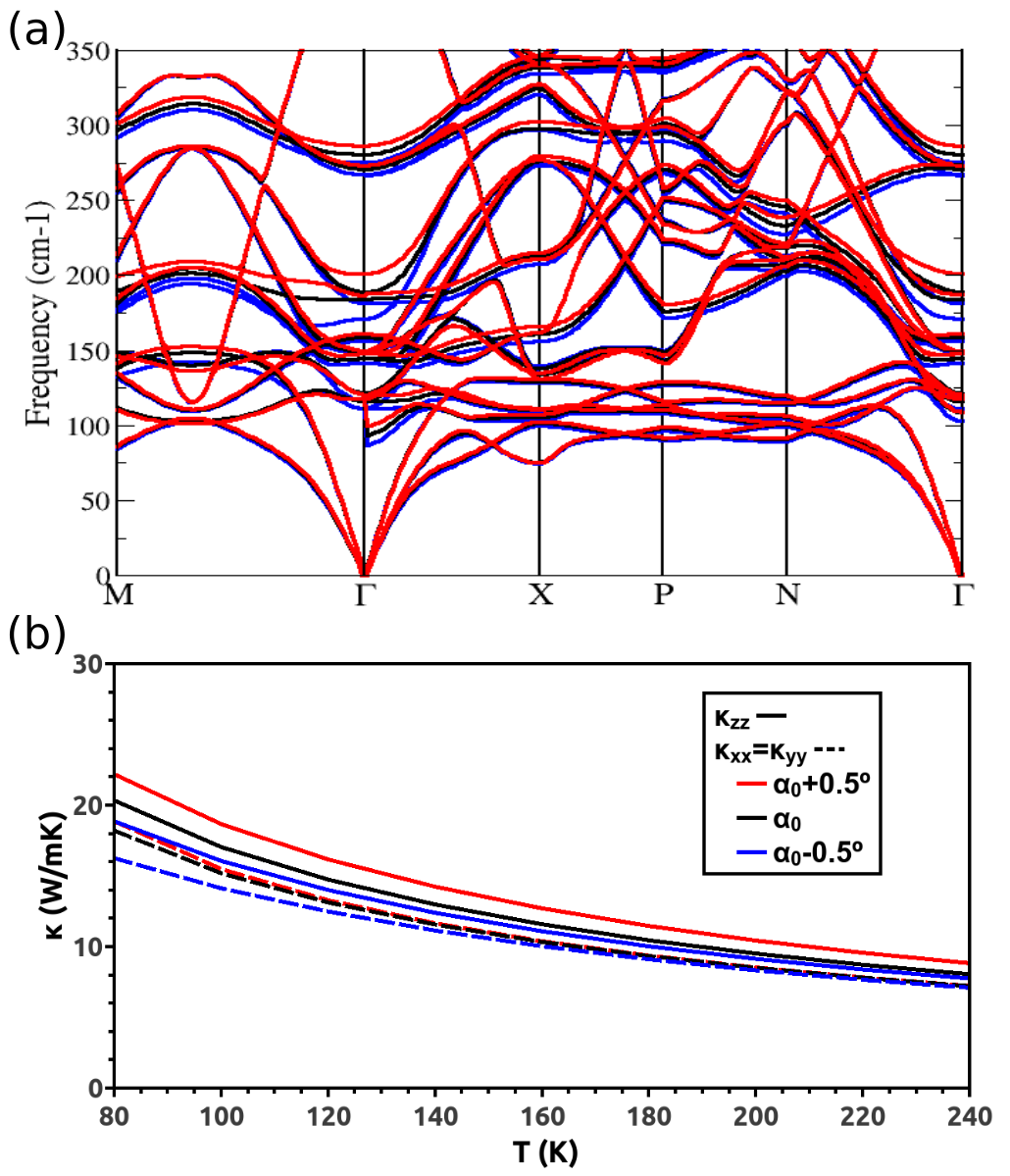}
     \caption{Results for the variation of the octahedral rotation angle in the $I4/mcm$ phase. Unit cell volume and $c/a$ ratio are kept constant. (a) Phonon band structure for three different angles, in red the largest and in blue the smallest, $\alpha_{0}$ is the optimized rotation angle. An increase in the angle increases the frequency of the polar modes. (b) Thermal conductivity as a function of temperature. Increasing the angle raises the thermal conductivity.}\label{kappa_angle}
\end{figure}

Figure \ref{kappa_ac}a shows the phonon dispersion for three different $c/a$ ratios. We can see that a reduction of the $c/a$ ratio increases the dispersion of the acoustic bands, thus increasing the group velocity of the compound  (Fig. \ref{dft_vel}b). We can also see that the z-axis polar mode softens when the $c/a$ ratio is increased. Moreover, the energy gap between this polar mode and the other doubly degenerate polar mode increases with $c/a$ (see Fig. \ref{scheme_polar}). Figure \ref{kappa_ac}b shows the increase of the thermal conductivity when $c/a$ is reduced. We observe that a decrease of $1\%$ in the $c/a$ ratio produces an increase of around $14\%$ on the thermal conductivity. The effect of the $c/a$ ratio on the thermal conductivity is relatively smaller compared to the effect of the  unit cell volume. The reason for this is that one polar mode softens and the doubly degenerate ones raise their frequency in the case of the $c/a$ distortion, while the three polar modes soften when modifying the volume (see Fig. \ref{scheme_polar}), increasing the phonon-phonon scattering and having a larger effect on $\kappa$.

Figure \ref{kappa_angle}a shows the phonon dispersion for three different octahedral rotation angles $\alpha$. We can see that increasing $\alpha$ has not a big effect on the dispersion of the acoustic bands, compared to the aforementioned volume distortions or varying the $c/a$ ratio. Mean group velocities of the acoustic modes increase with the angle, but barely compared to the other distortions. We can also see that the z-axis polar mode softens when $\alpha$ is reduced (see Fig. \ref{scheme_polar}). Figure \ref{kappa_angle}b shows the increase of the thermal conductivity when $\alpha$ is increased. This can be understood in terms of the evolution of the weighted phase space with the octahedral angle. Its increase causes a reduction of the peak. We observe that an increase of  $1^{\circ}$ in $\alpha$ produces an increase of around $14\%$ on $\kappa_{zz}$ while only $3\%$ on $\kappa_{xx}=\kappa_{yy}$. The effect of varying $\alpha$ on the thermal conductivity is lower compared to the effect of the two other distortions we have considered and analyzed. The main reason for this is that the acoustic bands are barely modified by $\alpha$, since there is no change in the lattice parameters.

The competition between the antiferrodistortive (octahedral rotations) and ferroelectric phases was already studied in literature, by suppressing the rotations one can in principle achieve a ferroelectric phase \cite{PhysRevB.62.13942, PhysRevB.82.174119, Benedek2013}. This could be achieved via strain engineering since different octahedral rotation patterns and angles can be obtained \cite{PhysRevB.95.165138}. We have shown here the coupling between the octahedral rotations and the polar mode responsible for ferrolectricity and also the effect that this has on the thermal conductivity of STO.

\section{Summary and conclusions} 

In this article we have analyzed, using a combined set of aiMD and DFT methods, the thermal conductivity of STO. We have studied the main sources of thermal transport and how these are modified when a distortion is introduced in the structure. We have analyzed both the high temperature cubic phase and the low temperature tetragonal one. 

In the cubic phase we have computed the phonon lineshape. This allowed us to identify the acoustic bands and the low-energy triply degenerate polar mode as the main heat carriers. The thermal conductivity that we have obtained is in good agreement with experiment. A reduction of the unit cell volume was found to increase the lattice thermal conductivity. This is associated to the increase of the acoustic bands' dispersion (group velocity) and also to a shift to higher energies of the polar mode, which produces a reduction of the scattering rate. The opposite behavior is found when the unit cell volume is increased. The degeneracy of the polar modes is broken if a tetragonal distortion is applied to that cubic phase. Due to this, the softening of the c axis polar mode and the change in the acoustic bands' dispersion produce a decrease in the thermal conductivity.

In the low-temperature phase we performed independently distortions of the unit cell volume, $c/a$  ratio and octahedral rotation angle. We found, as in the high temperature phase, that the acoustic bands and the polar modes control the variations of the thermal conductivity. Again, a reduction of the unit cell volume increases the acoustic bands' dispersion and raises the polar modes frequency, thus increasing the thermal conductivity. The reduction of the $c/a$ ratio (i.e., approaching a cubic structure) was found to increase the acoustic bands' dispersion and hence the thermal conductivity. Finally, the octahedral rotation angle was found to be strongly coupled to this phonon-active polar modes. We report that an increase of this angle raises the energy of the polar modes, decreasing the scattering rate and increasing the thermal conductivity. 

It is important to note that, in particular in the cubic phase, the polar modes have higher frequency than the rotational modes associated to the octahedral rotations, but these occur mainly at the zone boundary and hence have a smaller influence on the thermal conductivity (have less ability to provide scattering phase space for the acoustic modes). Also, the polar modes are flatter than the rotational modes. All this is relevant to understand why the polar modes so strongly influence the thermal conductivity properties in STO.

The study that we have performed here for STO could be extended and used to understand the thermal conductivity of similar perovskite-based oxides. The effect that the parameters we have explored:  unit cell volume, $c/a$ ratio and octahedral rotation angle, cause on the thermal conductivity provide a guide to design new experiments and nanostructures based on a wide variety of oxides.

\section*{ACKNOWLEDGEMENTS}
This work is supported by the MINECO of Spain through the projects MAT2016-80762-R and PGC2018-101334-B-C21. A.O.F. thanks MECD for the financial support received through the FPU grant FPU16/02572. We made use of the facilities provided by the Galician Supercomputing Center (CESGA). Work at the University of Missouri was supported by the U.S. Department of Energy, Award Number DE-SC0014607.

\end{document}